# Co-Translational mRNA Decay in Plants: Recent advances and future directions

Jean-Marc Deragon[1,2], Rémy Merret[3*]

[1] CNRS-LGDP UMR 5096, 58 avenue Paul Alduy, 66860 Perpignan, France

[2] Université de Perpignan Via Domitia, LGDP-UMR5096, 58 avenue Paul Alduy, 66860 Perpignan, France

[3] Institut de biologie moléculaire des plantes, CNRS, Université de Strasbourg, Strasbourg, France

[*]Corresponding author: remy.merret@cnrs.fr

Jean-Marc Deragon, ORCID ID, 0000-0002-2476-4932

Rémy Merret (ORCID ID, 0000-0002-3790-1115)

**Abstract**

Tight regulation of messenger RNA (mRNA) stability is essential to ensure accurate gene expression in response to developmental and environmental cues. mRNA stability is controlled by mRNA decay pathways, which have traditionally been proposed to occur independently of translation. However, the recent discovery of a co-translational mRNA decay pathway (also known as CTRD) reveals that mRNA translation and decay can be coupled. While being translated, a mRNA can be targeted for degradation. This pathway was first described in yeast and rapidly identified in several plant species. This review explores recent advances in our understanding of CTRD in plants, emphasizing its regulation and its importance for development and stress response. The different metrics used to assess CTRD activity are also presented. Furthermore, this review outlines future directions to explore the importance of mRNA decay in maintaining mRNA homeostasis in plants.

**Highlights**

Co-translational mRNA decay is a highly conserved mechanism in plants and plays an important role in mRNA homeostasis for proper development and stress response.

**Introduction**

Controlling the steady-state level of messenger RNA (mRNA) is a prerequisite for regulating gene expression in living organisms. The steady state of mRNAs is determined by the balance between their transcription and degradation. mRNA degradation is an evolutionarily conserved mechanism that is highly regulated in all organisms. After several rounds of translation, mRNAs undergo degradation, which is initiated by the removal of the poly(A) tail through a deadenylation process (Nagarajan *et al.*, 2013; Eisen *et al.*, 2020). Deadenylated mRNAs are then targeted for degradation via a 3′-5′ or a 5′-3′ pathway. The 3′-5′ mRNA decay pathway is mediated by the exosome complex or by the exoribonuclease DIS3L2 (called SOV in Arabidopsis) (Parker and Song, 2004; Zhang *et al.*, 2010; Malecki *et al.*, 2013). The 5′-3′ mRNA decay pathway occurs after a decapping step followed by a degradation by the exoribonuclease XRN1 or XRN4 in plants (Jones *et al.*, 2012; Nagarajan *et al.*, 2013). In Arabidopsis, defects in mRNA decay cause various phenotypes, such as post-embryonic lethality in plants defective in decapping (Xu *et al.*, 2006), or growth defects and stress sensitivity in plants defective in 5′-3′ degradation (Potuschak *et al.*, 2006; Merret *et al.*, 2013; Wawer *et al.*, 2018; Windels and Bucher, 2018; Nagarajan *et al.*, 2019; Kawa *et al.*, 2020; Carpentier *et al.*, 2024).

It has long been proposed that translation and mRNA decay are mutually exclusive, with mRNA decay starting only after the release of the last translating ribosome. However, the discovery of mRNA decay intermediates in polysomes changed our view of the decay mechanism (Hu *et al.*, 2009). While being translated, a mRNA can be targeted for degradation by a decay pathway called co-translational mRNA decay (CTRD). This mechanism appears to be evolutionarily conserved, as it has been characterized in several organisms ranging from bacteria to plants (Pelechano *et al.*, 2015; Yu *et al.*, 2016; Guo *et al.*, 2023; Huch *et al.*, 2023; Zhang *et al.*, 2023; Stevens *et al.*, 2024).

The development of high-throughput sequencing strategies has revealed the importance of this pathway in plants, both during development and in response to stress. Several factors have also been proposed to regulate CTRD in plants. In this review, we explore recent advances in the characterization of CTRD in plants. We also summarize the different metrics that have been developed by several groups to assess CTRD activity. Finally, we discuss the future directions to dissect the molecular mechanisms that trigger CTRD in plants.

**From Splint-Ligation PCR to 5'P Sequencing**

The first evidence suggesting that mRNA decay can occur co-translationally was the identification of decay factors such as UPF1 in polysomes (Atkin *et al.*, 1995, 1997; Mangus and Jacobson, 1999). These observations were subsequently confirmed by the identification of uncapped mRNAs in yeast polysomes (Hu et al., 2009) using a splint-ligation assay (initially developed by Moore and Query, 2000). In this assay, the 5'-monophosphate terminus of uncapped mRNAs is ligated to an RNA adaptor using a complementary splint DNA oligonucleotide that bridges the 5'-end of the transcript of interest to the 3'-end of the adaptor. The ligated product can then be reverse transcribed, amplified and revealed on agarose gel (Hu *et al.*, 2009). This approach, performed on RNA isolated from polysome fractions, showed that mRNA decapping can occur directly in polysomes and was the first evidence for the existence of co-translational mRNA decay. A few years later, the existence of such a pathway in plants was demonstrated using a similar approach in Arabidopsis (Merret *et al.*, 2015). However, the splint-ligation PCR assay can only be applied to a few candidate transcripts. Subsequently, the development of high-throughput degradome sequencing approaches has revealed that this process globally shapes the entire transcriptome in many organisms (Addo-Quaye *et al.*, 2008; Pelechano *et al.*, 2015; Yu *et al.*, 2016; Ibrahim *et al.*, 2018; Tuck *et al.*, 2020; Guo *et al.*, 2023; Huch *et al.*, 2023; Stevens *et al.*, 2024). The various degradome sequencing approaches, known as genome-wide mapping of uncapped and cleaved transcripts (GMUCT), parallel analysis of RNA ends (PARE-seq), Akron5-Seq, degradome-seq or 5'Pseq, all capture and sequence the mRNA degradation intermediates containing a 5'-monophosphate terminus. (Addo-Quaye *et al.*, 2008; German *et al.*, 2008; Willmann *et al.*, 2014; Pelechano *et al.*, 2015; Ibrahim *et al.*, 2018; Carpentier *et al.*, 2021; Chen *et al.*, 2025). In this review, the term "5'Pseq" will be used to refer to all of these approaches. The first genome-wide description of co-translational mRNA decay was performed in yeast (Pelechano *et al.*, 2015). By analyzing 5'Pseq data, the authors observed that mRNA decay intermediates are heterogeneously distributed along gene bodies, but this distribution appears to be periodic within coding regions, with a clear 3 nucleotide periodicity suggesting a link to a translational process. When the analysis is focused on the stop codons, a clear 3-nt periodicity is observed with a large accumulation of 5'P reads 17 nt upstream of the stop codons. This is exactly the mRNA length covered by a ribosome stopping translation at the stop codon (Figure 1A). The 3-nt periodicity and the accumulation of 17 nt reads upstream of the stop codons could best be explained if the exoribonuclease XRN1 degrades mRNAs undergoing translation, following the last translating ribosome codon by codon (Figure 1A). In this way, XRN1 degradation is totally dictated by ribosome dynamics (Pelechano *et al.*, 2015). Since the termination step is a slow process, the accumulation of 5'P reads 17 nt before stop codons is higher than other positions along the CDS. This mechanism has also been confirmed in plants where XRN4 follows the last translating ribosome (Yu *et al.*, 2016). However, the accumulation of 5'P reads before the stop codons is slightly different, with the accumulation at 16 and 17 nt, suggesting that the region occupied on mRNAs by ribosomes stalled at the termination step is different in plants compared to yeast (Yu *et al.*, 2016; Carpentier *et al.*, 2020, 2024; Nersisyan *et al.*, 2020).

**Metrics used to assess Co-Translational mRNA Decay activity**

Since the development of 5'Pseq approaches, several metrics have been developed to assess CTRD activity. We summarize in this section the different metrics used in the literature. Since XRN4/XRN1 follows the last ribosome moving along the mRNA one codon at a time, a 3-nt periodicity is expected in the 5'Pseq data (Figure 1A). To assess this 3-nt periodicity, Nersisyan et al., 2020 have adapted a Fast Fourier Transformation algorithm (FFT signal). A Fourier transformation converts a signal (here 5'P read-ends) to a frequency representation (Figure 1B). When applied to 5'Pseq data, a clear FFT signal is observed with a period of 3. In addition to the 3-nt periodicity, the frame corresponding to the first nucleotide protected by the 5' boundary of the ribosome is expected to be enriched in the 5'Pseq data (Figure 1C). In all species analyzed so far, a clear enrichment of the second (F1) nucleotide of each codon is observed compared to the first (F0) and the third (F2) nucleotides (Figure 1C). For each frame, the relative preference compared to the other two frames can be calculated as the Frame Protection Index (FPI). This calculation can be performed on a gene-by-gene basis to determine transcripts in frame with co-translational mRNA decay. In Arabidopsis, while F1 is the most enriched frame, a slight enrichment of F2 compared to F0 has also been observed in some cases (Yu *et al.*, 2016; Guo and Gregory, 2023). To measure the intensity of co-translational decay in different contexts (e.g. in mutants or under stress conditions), several metrics can be used. Since the strong accumulation of 5'P reads at position 16/17nt before the stop codons is used as a hallmark of active CTRD (Pelechano *et al.*, 2015; Yu *et al.*, 2016), a metagene analysis around stop codons can be used to globally assess CTRD activity (Figure 1D). Indeed, a drastic reduction of this peak is observed in an *xrn4* mutant, in which CTRD is highly abolished (Yu *et al.*, 2016; Carpentier *et al.*, 2020, 2024). This metagene analysis was used, for example, to evaluate the role of DXO1 and FRY1 in CTRD regulation (Carpentier *et al.*, 2025) or to assess CTRD activity across Arabidopsis seedling development, during heat stress or during tomato circadian rhythm (Carpentier *et al.*, 2020; Zhang *et al.*, 2023; Dannfald *et al.*, 2025). Since metagene analysis can bias the signal towards highly abundant transcripts, a gene-by-gene analysis can be performed using the Terminational Stalled Index (TSI, Figure 1E). Specifically, the TSI is calculated by dividing the number of 5'P reads that accumulate 16 and 17 nt upstream of the first nucleotide of the stop codon by the average number of 5'P reads that accumulate within a 100 nt flanking region codon (Guo *et al.*, 2023). Since this index indicates the level of 5'P reads at the translation termination site, a high TSI value indicates high CTRD activity for the analyzed transcript. Another index has also been proposed called CRI (for Co-translational RNA decay Index, (Yu *et al.*, 2016)). This index is similar to the FPI but considered F1 and F2 as protected frames to take into account the slight enrichment of F2 compared to F0 observed in some cases (Yu *et al.*, 2016). Since, CTRD dynamically tracks ribosomes, the metagene analysis and these indexes can be applied to all codons. Such an analysis reveals for example that ribosome pausing can occur at some codon pairs or triplets in Arabidopsi*s* (Nersisyan *et al.*, 2020). To facilitate 5'Pseq analysis, a pipeline called FIVEPSEQ has been developed (Nersisyan *et al.*, 2020). FIVEPSEQ uses mapping files (bam files) to generate most of the metrics presented in this section and can be useful for analyzing CTRD under various conditions. CTRD has also been used to measure translation efficiency (Carpentier *et al.*, 2020; Dannfald *et al.*, 2025). Historically, translation efficiency has been measured as the ratio of polysome RNA-seq counts to total RNA-seq counts and more recently by the RiboSeq approach (Kudla and Karginov, 2016; Vélez-Bermúdez *et al.*, 2023). However, these two approaches cannot discriminate between capped translated mRNAs and uncapped translated mRNAs in a CTRD process. Normalizing reads found in polysomes to reads captured in 5'Pseq, is another way to calculate translation efficiency that correlates well with protein production (Carpentier *et al.*, 2020; Dannfald *et al.*, 2025). As the majority of 5'P reads corresponds to CTRD products, the sum of 5'P reads can be used to determine the proportion of uncapped mRNAs found in polysomes. By dividing polysomal RNAseq counts by 5'Pseq counts, an assessment of translation efficiency can be obtained (Carpentier *et al.*, 2020; Dannfald *et al.*, 2025). Other 5'P decay events such as endonucleolytic cleavage, Exon Junction Complex (EJC) footprint, cap site or miRNA slicing site can also be analyzed with 5'Pseq data and require specific bioinformatic pipelines (Hou *et al.*, 2016; Lee *et al.*, 2020; Han *et al.*, 2023; Nagarajan *et al.*, 2023; Pouclet *et al.*, 2024).

**Actors involved in CTRD regulation**

Because CTRD is a 5' to 3' decay mechanism, XRN4, the main cytoplasmic 5'-3' exoribonuclease in Arabidopsis, has been proposed to be the major enzyme involved in this pathway (Box 1A) (Yu *et al.*, 2016). Recently, the importance of CTRD in mRNA stability was tested using a transgenic line expressing an XRN4 protein specifically impaired in CTRD, called XRN4ΔCTRD (Box 1B) (Carpentier *et al.*, 2024). It appears that CTRD is a key determinant of mRNA stability and thousands of transcripts are exclusively targeted by this pathway. However, a slight CTRD activity remains in an *xrn4* mutant suggesting that other ribonucleases may be involved (Carpentier *et al.*, 2020, 2024). Recently, DXO1, another 5'-3' exoribonuclease in Arabidopsis, has also been proposed to be involved in CTRD (Carpentier *et al.*, 2025). DXO1 has a deNADding activity and triggers 5'-3' RNA decay for RNA harboring a $NAD^+$ cap (Kwasnik *et al.*, 2019). Using a catalytically inactive DXO1, the authors demonstrated that CTRD is impaired in this mutant, but to a lesser extent than in *xrn4*, probably by targeting a subset of transcripts harboring a $NAD^+$ cap (Carpentier *et al.*, 2025). The 3'-phosphoadenosine 5'-phosphate (PAP) is a potent inhibitor of 5'-3' mRNA decay in plants (Gy *et al.*, 2007). Under normal conditions, this metabolite is hydrolyzed to 5'AMP and Pi by the FIERY1 (FRY1) phosphatase (Gil-Mascarell *et al.*, 1999). CTRD activity is also sensitive to PAP as its activity is drastically reduced in a *fry1* mutant, and exogenous PAP treatment affects XRN4 and DXO1 association with polysomes and stabilizes CTRD targets (Han *et al.*, 2023; Carpentier *et al.*, 2025). The other exoribonucleases XRN2 and XRN3 do not appear to be involved in CTRD regulation (Han *et al.*, 2023). Cofactors for CTRD have also been described. The LA-related protein LARP1 is involved in the targeting of XRN4 to polysomes (Merret *et al.*, 2013). In a *larp1* knockout mutant, XRN4 accumulation to polysomes is abolished, leading to the stabilization of several transcripts (Merret *et al.*, 2013). PELOTA and HBS1 (two ribosome rescue factors) have been proposed to suppress CTRD activity, suggesting that translation termination regulation regulates CTRD activity (Guo and Gregory, 2023). In addition, the CAP BINDING PROTEIN 80 (CBP80), the largest subunit of the nuclear mRNA cap binding complex, is also involved in CTRD activity, as a moderate decrease in CTRD activity is observed in a *cbp80* knockout mutant (Yu *et al.*, 2016). Interestingly CBP80 contributes to the initial rounds of translation suggesting a link between translation regulation and CTRD activity (Kim *et al.*, 2009).

**Physiological importance of CTRD in plant development and stress response**

The first evidence of the physiological importance of CTRD in plants was demonstrated during heat stress (Box 1C) (Merret *et al.*, 2015). A short period of heat stress (15 minutes at 38°C) induces a general slowing of translation. In particular, ribosome pausing occurs on mRNAs encoding hydrophobic peptides, triggering XRN4 recruitment to polysomes and inducing CTRD (Merret *et al.*, 2015). More recently, the importance of CTRD for thermotolerance has been proposed (Box 1C) (Dannfald *et al.*, 2025). A priming event (60 minutes at 37°C) allows the induction of the Unfolded Protein Response (UPR) pathway during a subsequent acute heat stress (30 minutes at 44°C). In the absence of priming, UPR pathway transcripts are targeted by CTRD (Dannfald *et al.*, 2025). The importance of CTRD during diurnal rhythm was also tested in tomato (Box 1D) (Zhang *et al.*, 2023). In a time course experiment over one diel cycle, the authors showed that the mRNA decay is highly coordinated during light/dark oscillations. In particular, diurnally regulated mRNAs are targeted by CTRD to enable rapid mRNA turnover (Zhang *et al.*, 2023). Interestingly, an increase in CTRD during intense light stress and recovery has also been proposed (Crisp *et al.*, 2017). CTRD also appears to be important during plant development, as CTRD targets vary across seedlings development in Arabidopsis (Box 1E) (Carpentier *et al.*, 2020). Recently, the role of CTRD has also been tested at the organ level (Carpentier *et al.*, 2024). The absence of XRN4-mediated CTRD in Arabidopsis shoot and root induces a growth defect. CTRD has also been identified in flowers, suggesting that this pathway is a key hub in maintaining mRNA homeostasis for proper organ development (Yu *et al.*, 2016). CTRD regulation has also been proposed to be essential for seed germination (Box 1F) (Guo and Gregory, 2023). By analyzing 5'Pseq data generated from multiple angiosperm transcriptomes, CTRD features appear to be evolutionarily conserved (Box 1G) (Guo *et al.*, 2023). Until now, this pathway has been described in *Arabidopsis thaliana, Solanum lycopersicum, Oryza sativa, Glycine max, Zea mays, Setaria viridis, Sorghum bicolor, Brachypodium distachyon, Medicago truncatula, and Phaseolus vulgaris* (Hou *et al.*, 2016; Guo *et al.*, 2023).

**Conclusions and Future Directions**

Since its discovery in plants in 2015, important progress has been made in uncovering and deciphering the co-translational mRNA decay pathway. However, its characterization is only just beginning, and several questions remain. One of the key steps for triggering mRNA decay is a reduction in the size of the mRNA poly(A) tail through a deadenylation process. In yeast and mammals, the deadenylation complex, CCR4-Not, has been implicated as a key element in coupling ribosome slowing-down and mRNA decay (Buschauer *et al.*, 2020; Zhu *et al.*, 2025). The requirement for deadenylation and the role of the CCR4-Not complex in triggering CTRD in plants is still an open question. The recent development of long-read sequencing dedicated to poly(A) tail length (Jia *et al.*, 2022) should help to better understand the role of deadenylation in triggering CTRD. Indeed, using direct RNA sequencing, it has recently been proposed that stress-induced RNA decay may be independent of deadenylation in human (Dar *et al.*, 2024). In yeast, the RNA helicase DHH1, has also been proposed to sense ribosome slowing down to trigger CTRD (Radhakrishnan *et al.*, 2016). In Arabidopsis, 3 DHH proteins exist and have been shown to be important for mRNA decay particularly for stress-response mRNAs (Chantarachot *et al.*, 2020). However, their roles in CTRD have not yet been demonstrated (Chantarachot *et al.*, 2020; Merret and Bousquet-Antonelli, 2020). Recently, the 3'-5' mRNA decay pathway mediated by the exosome complex has also been proposed to occur co-translationally in humans (Kögel *et al.*, 2024). A supercomplex composed of the exosome and the ribosome serves as a platform for ribosome collision to enable rapid mRNA quality control. Interestingly, in Arabidopsis, HBS1 (also called SKI7), a member of the exosome complex has been implicated in CTRD regulation (Guo and Gregory, 2023). In several organisms, the ribosome appears to be a hub for controlling mRNA homeostasis (recently reviewed in Müller et al., 2025). Emerging NGS sequencing approaches and cryo-EM structural analysis in plants promise new insights into the interplay between translation and mRNA decay. Future research on these aspects will be important to improve our understanding of the mechanisms involved in maintaining mRNA homeostasis in plants.

**Acknowledgements**
This study is set within the framework of the "Laboratoires d'Excellences (LABEX)" TULIP (ANR-10-LABX-41) and of the "École Universitaire de Recherche (EUR)" TULIP-GS (ANR-18-EURE-0019)

**Conflict of interest**

None

**Funding**

The work of RM is supported by the Agence Nationale de la Recherche (ANR-21-CE20-0003). The work of JMD is supported by the Agence Nationale de la Recherche (ANR-22-CE20-0023) and by a New Frontier grant of the LABEX TULIP (ANR-10-LABX-41).

**Figure 1: Metrics used to assess co-translational mRNA decay activity.** A. Schematic representation of the co-translational mRNA decay pathway. XRN4 follows the last translating ribosome codon by codon. The first nucleotide protected by the 5' boundary of the ribosome harbors a 5' monophosphate extremity that can be captured by 5'Pseq. B. The Fast Fourier Transformation (FFT) signal calculates the periodicity frequency and can be used to illustrate a 3-nt periodicity in 5'Pseq data. C. The Frame Preference Index (PFI) can be used to determine transcripts in frame with CTRD. An enrichment in frame 1 (F1) is expected. D. A metagene analysis of 5'P reads accumulation around stop codons can be performed to assess global CTRD activity. A 5'P reads accumulation 17 nucleotides (16/17nt in plants) before stop codons is expected as this position corresponds to the ribosome reaching the stop codon in the A site. The Terminational Stalling Index (TSI) or the Co-translational RNA Decay Index (CRI) can be used to assess CTRD activity at the level of individual transcripts.

**(a) Representation of the co-translational mRNA decay mechanism**

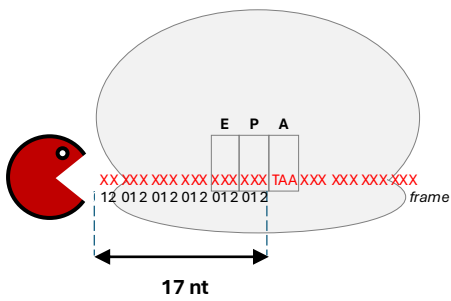
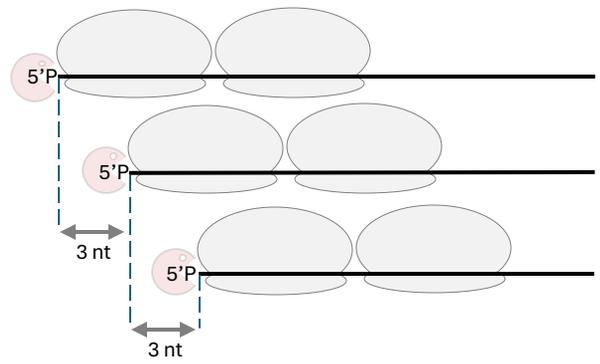

**(b) 3-nt periodicity (FFT signal)**

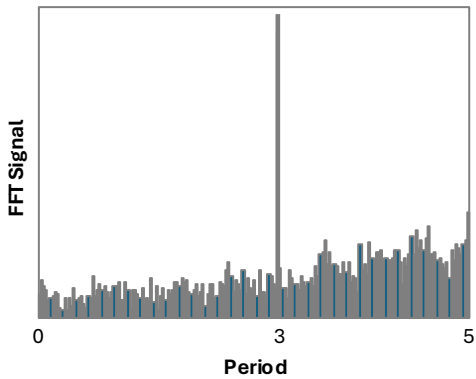

**(c) Frame Preference Index (FPI)**

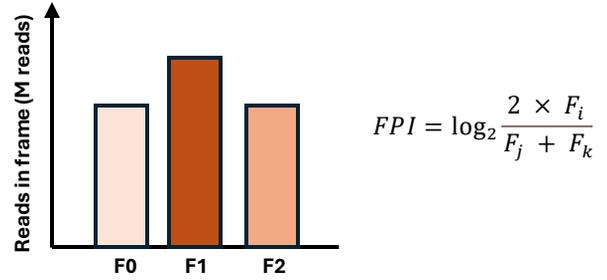

$$FPI = \log_2 \frac{2 \times F_i}{F_j + F_k}$$

**(d) Metagene Analysis around stop codons**

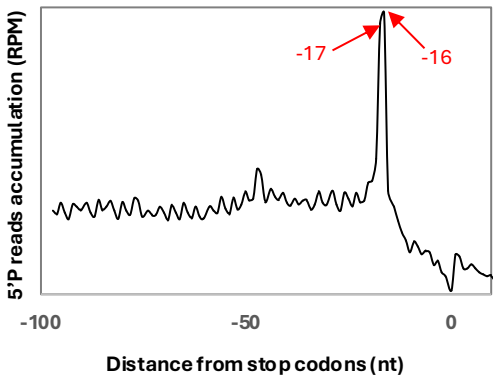

**(e) Individual co-translational decay activity**

**Terminational Stalling Index (TSI)**

$$TSI = \frac{\sum 5'P \text{ read ends } 16 \text{ and } 17 \text{ nt before stop codons}}{\text{Mean } 5'P \text{ read ends in a } 100 \text{ nt flanking region}}$$

**Co-translational RNA decay Index (CRI)**

$$CRI = \log_2 \frac{\frac{F_1 + F_2}{2}}{F_0}$$

**Figure 1 : Metrics used to assess co-translational mRNA decay activity.** A. Schematic representation of the co-translational mRNA decay pathway. XRN4 follows the last translating ribosome codon by codon. The first nucleotide protected by the 5' boundary of the ribosome harbors a 5' monophosphate extremity that can be captured by 5'Pseq. B. The Fast Fourier Transformation (FFT) signal calculates the periodicity frequency and can be used to illustrate a 3-nt periodicity in 5'Pseq data. C. The Frame Preference Index (PFI) can be used to determine transcripts in frame with CTRD. An enrichment in frame 1 (F1) is expected. D. A metagene analysis of 5'P reads accumulation around stop codons can be performed to assess global CTRD activity. A 5'P reads accumulation 17 nucleotides (16/17nt in plants) before stop codons is expected as this position corresponds to the ribosome reaching the stop codon in the A site. The Terminational Stalling Index (TSI) or the Co-translational RNA Decay Index (CRI) can be used to assess CTRD activity at the level of individual transcripts.

**Box 1. Key developments in understanding the role of co-translational mRNA decay in plant development and stress response**

(A) Yu et al. (2016) provided the first evidence that XRN4 is the main enzyme involved in CTRD in Arabidopsis.

(B) Carpentier *et al.* (2024) demonstrated that the CTRD pathway is the main 5′-3′ mRNA decay pathway in Arabidopsis. Using a specific XRN4 transgenic line impaired in CTRD (called XRN4ΔCTRD), the authors clearly defined the importance of CTRD in the general mRNA turnover.

(C) Merret et al. (2015) and Dannfald et al., (2025) illustrated the importance of CTRD in plant thermotolerance. A short period of heat stress (15 minutes at 38°C) induces the slowing of ribosomes, which triggers CTRD. During acute heat stress (15 minutes at 44°C), CTRD targets transcripts involved in the UPR response.

(D) Zhang et al. (2023) analyzed the importance of CTRD in the circadian rhythm in tomato and demonstrated that CTRD contributes to the rapid turnover of diurnal mRNAs.

(E) Carpentier et al. (2020) analyzed the dynamics of CTRD during Arabidopsis seedling development and demonstrated that CTRD targets vary across development to properly regulate protein production.

(F) Guo and Gregory (2023) identified two ribosome rescue factors involved in CTRD regulation with potential roles in seed germination.

(G) Guo et al. (2022) analyzed 5′Pseq data from 10 different angiosperms' transcriptomes and discovered that CTRD is likely conserved across plant lineages.